\documentclass{article}
\usepackage{spconf,amsfonts,amsmath,graphicx}
\usepackage{booktabs}
\usepackage{multirow}
\usepackage{CJKutf8}
\usepackage{makecell}

\title{Practice of the conformer enhanced AUDIO-VISUAL HUBERT on Mandarin and English}
\name{Xiaoming Ren,  Chao Li, Shenjian Wang, Biao Li}
\address{Beijing OPPO telecommunications corp., ltd., Beijing, China}
\begin{document}
\ninept
\maketitle
\begin{abstract}
Considering the bimodal nature of human speech perception, lips, and teeth movement has a pivotal role in automatic speech recognition. Benefiting from the correlated and noise-invariant visual information, audio-visual recognition systems enhance robustness in multiple scenarios. In previous work, audio-visual HuBERT appears to be the finest practice incorporating modality knowledge. This paper outlines a mixed methodology, named conformer enhanced AV-HuBERT, boosting the AV-HuBERT system's performance a step further. Compared with baseline AV-HuBERT, our method in the one-phase evaluation of clean and noisy conditions achieves 7\% and 16\%  relative WER reduction on the English AVSR benchmark dataset LRS3. Furthermore, we establish a novel 1000h Mandarin AVSR dataset CSTS. On top of the baseline AV-HuBERT, we exceed the WeNet ASR system by 14\% and 18\% relatively on MISP and CMLR by pre-training with this dataset. The conformer-enhanced AV-HuBERT we proposed brings 7\% on MISP and 6\% CER reduction on CMLR, compared with the baseline AV-HuBERT system.
\end{abstract}

%
\section{Introduction}
\label{sec:intro}
In recent years, automatic speech recognition (ASR) systems have seen considerable improvements with the help of innumerable neural networks and models \cite{Hinton2012DeepNN}, which reach or exceed mankind in several scenarios, especially low-noise, near-field situations \cite{Nguyen2021SuperHumanPI}. A considerable amount of literature has been published on end-to-end approaches \cite{Kim2017JointCB}\cite{Graves2012SequenceTW}\cite{Chan2015ListenAA}. Among those attempts, attention-based architectures seem to become prevailing and receive rave reviews, such as conformer \cite{Gulati2020ConformerCT}, which is able to learn local context and long-period connections when modeling sequences.  

Nevertheless, ASR performance degrades inevitably due to various external disturbances, considering the vast applicability range of speech recognition technology. However, lip movements highly correlated to human speech are not benefited from the correlated and noise-invariant visual information, audio-visual recognition systems (AVSR) enhance robustness in multiple scenarios. By capturing the delicate relationship between audio and visual information, the treasure behind would be exploited.

With the availability of well-designed end-to-end architectures, the help of a growing number of multimedia datasets \cite{Afouras2018LRS3TEDAL}\cite{Chung2018VoxCeleb2DS} and the fusion of visual and audio modalities \cite{Sterpu2018AttentionbasedAF}\cite{Sterpu2020HowTT}, AVSR has remarkably made significant progress in recent years. Audio-Visual Hidden Unit BERT (AV-HuBERT) \cite{Shi2022LearningAS}, a self-supervised AVSR framework with a pre-training and fine-tuning stage, seems to bring AVSR performance to a new level.  

In this paper, we present a mixed methodology to boost AV-HuBERT system CER performance further. Firstly, by adopting the 80-dim filterbank feature, audio knowledge is more thoroughly and effectively captured. Secondly, a modified version of ResNet Encoder \cite{Stafylakis2017CombiningRN} which is pre-trained deliberately with an abundant number of multimedia resources, is employed to extract visual information. Then, we innovatively suggest a modality fusion method with a gating mechanism. Audio along with visual information is adopted in the design of the fusion gate, balancing audio knowledge going through the system. Finally, we investigate the usage of the conformer instead of the transformer, as mentioned above, which reduces deletion error in speech recognition and thus helps our system evolve further. 
We also establish a Mandarin multimodal dataset containing 1000h Mandarin audio and visual data. A Mandarin model based on AV-HuBERT is pre-trained using this dataset, which outperforms WeNet by 14\% and 18\% on MISP and CMLR after fine-tuning.
Furthermore, we propose the enhanced AV-HuBERT, which boosts baseline AV-HuBERT performance by a relative 7\% and 6\% on MISP and CMLR.

The rest of this paper is organized as follows. Section 2 reviews recent works in the AVSR domain. Section 3 profoundly presents our methodology. Section 4 introduces the dataset and how we conduct those experiments. The result and related analysis are presented in Section 5. Section 6 concludes and also discusses further work.

\section{Related Work}
\label{sec:format}
\noindent\textbf{AV-HuBERT.}
It has previously been observed that modern novel neural networks are craving hand-labeled data for training since they are fully supervised. Most AVSR systems were no exception. Encouragingly, \cite{Shi2022LearningAS}\cite{Shi2022RobustSA} proposed a robust self-supervised AVSR framework AV-HuBERT, capturing modality information from human speech and lip movement at the same time and achieved state-of-the-art on AVSR benchmark dataset LRS3 \cite{Afouras2018LRS3TEDAL}. Feature clustering and masked prediction are two impressive aspects of the pre-training stage. In this paper, our work is built on top of AV-HuBERT, continuing to evolve and grow to a more robust AVSR system.

\noindent\textbf{Conformer.}
Attention-based transformer plays a critical role and seemed to become prevailing. Nonetheless, its variant conformer, which can capture local context through convolution layers and long-term relationships, is even better. In this paper, we adopt a conformer encoder instead of the transformer in the pre-training stage, hoping to learn the nuanced correlation and local information. 

Standard conformer block \cite{Gulati2020ConformerCT} is composed of four modules, including a feed-forward module, a self-attention module, a convolution module, and a second feed-forward module. The two feed-forward modules sandwich the multi-headed self-attention module and the convolution module.  

Formally, for input \(\mathbf{x}_{i}\)
to a conformer block \(i\), the output \(\mathbf{y}_{i}\) of the block is computed as below:

\begin{align}
\widetilde{\mathbf{x}}_{i} &= \mathrm{LN}\left(\mathbf{x}_{i} + \frac{1}{2}\mathrm{FFN}(\mathbf{x}_{i})\right) \\
{\mathbf{x}_{i}}' &= \mathrm{LN}(\widetilde{\mathbf{x}}_{i} + \mathrm{MHSA}(\widetilde{\mathbf{x}}_{i})) \\
{\mathbf{x}_{i}}'' &= \mathrm{LN}({\mathbf{x}_{i}}' + \mathrm{Conv}({\mathbf{x}_{i}}')) \\
{\mathbf{y}_{i}} &= \mathrm{LN}\left({\mathbf{x}_{i}}'' + \frac{1}{2}\mathrm{FFN}({\mathbf{x}_{i}}'')\right)
\end{align}

%
where FFN refers to the feed-forward module, MHSA refers to the multi-head self-attention module, Conv refers to the
convolution module and LN refers to the layer norm module.

\begin{figure}[t]
  \centering
  \includegraphics[width=\linewidth]{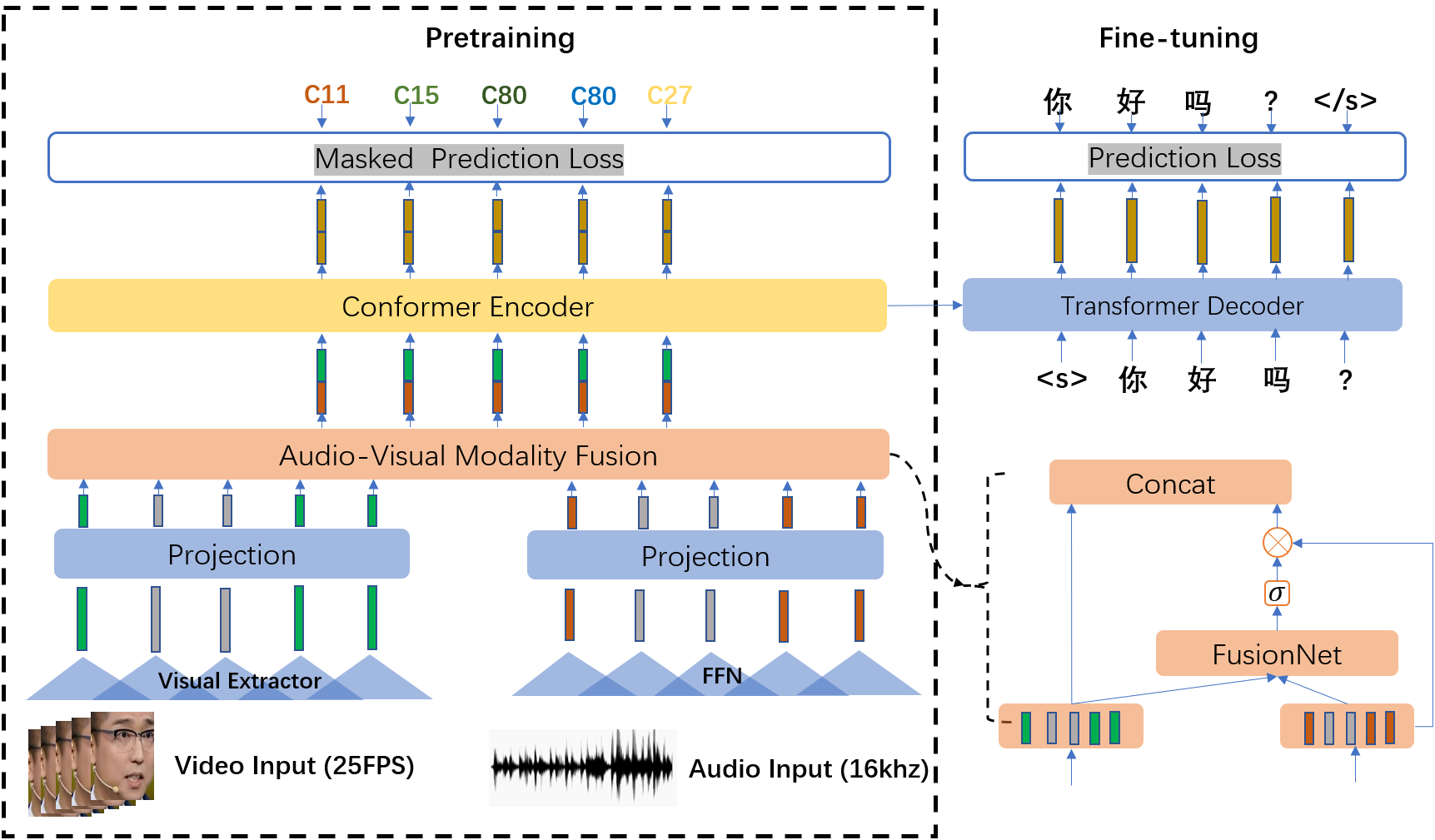}
  \caption{Attention information visualization between different layers and heads, indicates the diversity of information, and complementarity.}
  \label{fig:speech_production}
\end{figure}

\section{THE PROPOSED METHOD}
\label{ssec:subhead}

\subsection{Audio Feature}
As we repeat the AV-HuBERT research and reproduce the result on speech recognition,  it is interesting to note that in the previous experiments, the audio feature employed in the pre-training process is 26-dim filterbanks. Empirically, a higher dimension gives a better representation of knowledge in several demanding and complex situations. Hence, we adopt 80-dim filterbanks as many end-to-end systems do. Concerning this adjustment, the performance sees a stable improvement. Additionally, when carrying out the research, features computed from a 15ms window outperform those from a 25ms window. 
\subsection{Visual Feature Extractor}
As the filterbank feature is popular for the audio stream, the visual information is extracted with front-end 3D ResNet18, whose parameters are learned during the model training stage. This visual extractor seems to more or less rely on the quality and quantity of its data set, which tends to be vulnerable and ephemeral in certain cases. We employ light-weighted MobileNet-V2 \cite{Sandler2018MobileNetV2IR} instead of ResNet.

\subsection{Gated Fusion}
Enlightened by \cite{Yu2020AudioVisualRO}, a mixed modality fusion methodology is introduced to our enhanced AV-HuBERT system. The paper above demonstrates a comparison between straight concatenation of modality features and fusion with gating mechanism \cite{Dauphin2017LanguageMW}. We believe simple concatenation may lead to unreliable results when the visual flow is of low quality or not synchronized with audio. Hence, the audio and visual information are combined to control the audio information flow with the help of a GLU gate. At this point, the chosen audio knowledge is then incorporated with its visual part as the input of the pre-trained encoder. 
We incorporate this scheme into the AV-HuBERT model. The structure of fusion is shown in Fig.1. FusionNet output is \(\mathbf{m}_{t}\).

\begin{align}
\mathbf{m}_{t} &= (\mathrm{concat}(\mathbf{v}_t,\mathbf{a}_t))\ast \mathbf{U} + \mathbf{a} \\
\mathbf{h}_{t} &= (\mathbf{a}_t\ast \mathbf{W}+\mathbf{b} )\otimes  \sigma (\mathbf{m}_t\ast \mathbf{V}+\mathbf{c})
\end{align}


where \(\mathbf{U}\in \mathbb{R}^{2D\times D}\),\(\mathbf{W}\in \mathbb{R}^{D\times D}\),\(\mathbf{V}\in \mathbb{R}^{D\times D}\), \(\{\mathbf{a},\mathbf{b},\mathbf{c}\} \in \mathbb{R}^{D}\)  are the bias, \(\sigma\) refers to the sigmoid function, \(\otimes\) denotes the Hadamard product.

\subsection{Conformer Encoder}
Focusing back on the encoder in the original AV-Hubert architecture, one thing to note is that the transformer encoder employs convolution for its position encoding. We find it not compatible with conformer structure, and hence we employ the relative sinusoidal positional encoding scheme which proved to be an important technique from Transformer-XL \cite{Dai2019TransformerXLAL}. The relative positional encoding scheme supports the self-attention module on different input lengths, leading to robustness towards various utterance lengths. The effectiveness of this adjustment is demonstrated later. Moreover, blockformer \cite{Ren2022ImprovingMS} supplies the last output layer with adequate previous block output, which indicates the potential to be of benefit. Surprisingly, we also find that the conformer system reduces the reduction error in speech recognition, compared with the transformer system.

\begin{CJK*}{UTF8}{gbsn}
\begin{table}[t]\footnotesize
  \caption{\noindent{Composition of the CSTS dataset.} }
  \label{tab:word_styles2}
  \centering
  \begin{tabular}{|c|c|c|c|c|}
    \hline

    source & \makecell[c]{raw \\ hours} & \makecell[c]{processed \\ hours}  & utt & spks                 \\
    \hline
     CC Forum   & {215}    & 101 & 50670 & 791 \\
     Topics in Focus   & {243}    & 69 & 33141 & 3601  \\
     Yixi Lecture   & {39}    & 13 & 6043 & 91  \\
     Du Talk-Show   & {161}    & 117 & 59017 & 694  \\
     Legal Report   & {1997}    & 598 & 317154 & 14209  \\
     Rock\&Roast   & {73}    & 7 & 3451 & 69  \\
     The News Broadcasting   & {482}    & 82 & 38023 & 672  \\
     News Live Room & 298 & 69 & 33321          & 548        \\
      \hline
     Total & 3508 & 1056 & 540820 & 20675 \\

    \hline
  \end{tabular}
\end{table}
\end{CJK*}
\section{ Experimental setting}

Our experiments are based on four datasets, including:

\noindent\textbf{LRS3 .}
The dataset consists of over 400 hours of video, extracted
from 5594 TED and TEDx talks in English, downloaded from
YouTube. The dataset is organized into three sets: pre-train, train-val
and test. The first two overlap in terms of content but the last is
completely independent. It is the largest publicly available labeled audio-visual speech recognition dataset.

\noindent\textbf{CSTS.} We collect a 1000h unsupervised Chinese audio-visual dataset containing 200k individuals, which will release with the paper. Since it contains only one speaker at a time, we name this dataset Chinese Solo Talk Show (CSTS). The CSTS dataset is organized into three sets: pre-train, train-val, and test. 

The cropped face tracks are provided as .mp4 files. The audio tracks are provided in single-channel 16-bit 16kHz format.
Table 1 demonstrates the source, period, total sentence, and number of speakers. We collect speeches, interviews, and news programs from Chinese websites. After face detection, face clustering, face recognition, and VAD, the dataset is available and ready for use. 

\noindent\textbf{CMLR.} 
It is a large Chinese Mandarin Lip Reading dataset (CMLR), designed to facilitate research on visual speech recognition, sometimes also referred to as automatic lip reading. More than 100,000 spoken sentences from 11 speakers of the national news program “News Broadcast” are included, up to an estimated total length of 61+h. More than 3,000 Chinese characters and 20,000 phrases appear in CMLR.
This dataset includes train, dev, and test sets. Following regular data allocation, we use the train and dev set(up to 61h) for fine-tuning and the 17h test set for decoding.

\noindent\textbf{MISP.}
This dataset is from the Multimodal Information Based Speech Processing Challenge 2021 (MISP), which contains 100+h of audio-visual data. The dataset is divided into the near, middle, and far scenarios, and recorded with 30 rooms, and 248 participants condition. The near-field audio is recorded individually via high-fidelity hardware and the middle scenario video is synchronized with the audio. Thus, we use near-scenario audio along with middle-range video, excluding far scenarios.

Generally speaking, our experiments are consistent with AV-HuBERT \cite{Shi2022LearningAS}, including pre-training and fine-tuning stages. The pre-training stage learns feature representation through unsupervised audio and visual data, while fine-tuning uses labeled audio-visual pairs, according to the parameters of the labeled data.

\subsection{Pre-Training}
\noindent\textbf{Pseudo labels.}
We divide our experiments into the one-phase evaluation and the five-phase operation. We use 100 pseudo labels in one-phase evaluation, while the five-phase is the same as AV-HuBERT. 

In five-phase, we iteratively increase the label number from 100, 100, 500, 1000, to 2000 at last. Note that in the first phase, we only use MFCC-39 from speech frames, in the other phases the transformer encoder's output from the previous phase is the input for feature clustering. 

\noindent\textbf{Audio feature.}
Filterbanks are for feature extraction. We compare fbank26 with fbank80, window length of 25ms with 15ms. If not mentioned, we adopt fbank26+win25. 

\noindent\textbf{Visual feature.}
We experiment with two methods of extracting visual features: resnet refers to ResNet18; mobilenet refers to MobileNet-v2.

\noindent\textbf{Modality fusion.}
In feature fusion experiments, we compare simple concatenation with the gated fusion method. In Table 3 the GLU refers to the gated way.

\noindent\textbf{Backbone.}
Transformer vs conformer backbone is compared in this paper. The Transformer encoder shares the same structure with AV-HuBERT while the conformer encoder uses relative position encoding, stacking 12 conformer blocks. In detail, the encoder embedding dimension is set to 768, the layer drop is 0.05, and the remaining keep the same with the transformer. The total parameter in the transformer is 103M and 183M for the conformer.

We conduct one-phase evaluations on LRS3 with 433h pre-training and 30h fine-tuning. The training stage is based on 8 A100 GPUs with max tokens of 1000. It takes 64h, containing 400k steps.

The comparison with WeNet ASR on the Mandarin set is based on five-phase operations. Enhanced AV-HuBERT ablation studies are just one phase. Our Mandarin pre-training models are all built upon a 1000h CSTS dataset and fine-tuned on 100h MISP and 61h CMLR afterward. Our model is trained on 4 A100 GPUs with max tokens 2500, one batch size of data containing at most 100 seconds of audio per GPU. It takes around 32 hours to finish a phase, where 400k steps are. We do not use accumulated gradient, update freq is set to 1.

 

One thing to note is that, among all pre-training and fine-tuning stages, there is no extra noise added. Random noise is included in the decoding stage. The noise audio clips in the categories of ”natural”, ”music” and ”babble” are sampled from the MUSAN dataset \cite{Snyder2015MUSANAM}.

\subsection{Supervised Fine-Tuning and Decoding}

As for the fine-tuning stage, we use the attention-based sequence-to-sequence cross-entropy
loss \cite{Bahdanau2016EndtoendAL}. Two modalities are included in fine-tuning. The modeling unit here for LRS3 is unigram-based subword units \cite{Kudo2018SubwordRI}. The vocabulary size is 1000.

As for the Mandarin dataset, Chinese characters are employed as the modeling unit. Linguistic knowledge is learned by the seq2seq transformer decoder of its own accord. 
Parameters for Mandarin, as well as English datasets in the fine-tuning stage, are listed below.

No extra language model is added during inference. It takes 30k steps to fine-tune the six-layer transformer decoder where in the first 24k steps, parameters in the pre-trained model are frozen. In the remaining 6k steps, we unfreeze the parameters. Other settings are similar to AV-HuBERT.

Warming up is set to 10k and learning is 0.001, without accumulated gradient. Adam \cite{Kingma2015AdamAM} optimizer is used with \(\beta\)  = (0.9, 0.98).
We use 8 A100 GPUs in fine-tuning process, with the max token set to 1000, which means a batch size of at most 40 seconds. 

\begin{table}[t]\footnotesize
  \caption{\noindent{Ablation study of the enhanced AV-HuBERT on LRS3 with update freq 1 for first phase (WER:\%). The baseline is AV-HuBERT. All is composition of fbank80, win15, GLU and mobilenet.}}
  \label{tab:word_styles2}
  \centering
  \begin{tabular}{|c|c|c|c|c|c|}
    \hline
  \multirow{2}*{\textbf{Method}} &  \multicolumn{2}{|c|}{\textbf{Audio-only}} &  \multicolumn{2}{|c|}{\textbf{Audio-visual}} \\
   ~ & \textbf{CWER}& \textbf{NWER} & \textbf{CWER}& \textbf{NWER}   \\
    \hline
    baseline             & 23.8   & 87.52  & 15.88  & 46.97          \\
    baseline+fbank80     & /      & /      & 15.24  & 46.26          \\
    baseline+win15       & 28.65  & 82.74  & 15.66  & 44.43          \\
    baseline+GLU         & 21.07  & 71.42  & 16.31  & 49.59          \\
    baseline+mobilenet   & 21.84  & 69.25  & 16.36  & 48.15          \\
    baseline+all        & 19.81  & 68.70  & 14.34  & 43.95          \\

    \hline
  \end{tabular}
\end{table}

\begin{table}[t]\footnotesize
  \caption{\noindent{Ablation study of the enhanced AV-HuBERT on LRS3 with update freq 4 for first phase (WER:\%). The implication of baseline and all is the same as Table 2.} }
  \label{tab:word_styles2}
  \centering
  \begin{tabular}{|c|c|c|c|c|c|}
    \hline
  \multirow{2}*{\textbf{Method}} &  \multicolumn{2}{|c|}{\textbf{Audio-only}} &  \multicolumn{2}{|c|}{\textbf{Audio-visual}} \\
   ~ & \textbf{CWER}& \textbf{NWER} & \textbf{CWER}& \textbf{NWER}   \\
    \hline
    baseline             & 17.59  & 82.04  & 12.56  & 44.39          \\
    baseline+conformer   & 15.08  & 59.54  & 13.09  & 39.25          \\
    baseline+conformer+all         & 14.93  & 82.18  & 11.66  & 37.09          \\

    \hline
  \end{tabular}
\end{table}

\section{Experimental results \& Analysis}

In Table 2, every boosting method is verified on LRS3, where 'audio-only' refers to the case when the audio feature is employed for decoding. By 'audio-visual', we use both audio and visual features at the same time. CWER refers to CER in a clean environment and NWER is the WER with random noise in the decoding stage. We find that, in fbank80, win15 conditions, WER reduces for both clean and noisy cases. With GLU and mobilenet, audio-only WER receives a reduction in both cases. However, the word error rate grows with audio-visual in clean or noisy experiments.
We combine all the boosting methods and the result is shown in the last row. Clean WER drops up to 16.8\% and 9.7\% relatively with audio and audio-visual features respectively. Noisy WER drops relatively by 21.5\% and 6.4\% respectively.

\begin{table}[t]\footnotesize
  \caption{\noindent{Ablation study of the enhanced AV-HuBERT pre-training on CSTS for the first phase and fine-tuning on MISP (CER:\%).}  }
  \label{tab:word_styles2}
  \centering
  \begin{tabular}{|c|c|c|c|}
    \hline
    \textbf{Backbone} & \textbf{Method} & \textbf{CER} \\
    \hline
    \multirow{5}*{transformer}     & fbank26+win25+concat+resnet & 18.93                  \\
    ~    & \textbf{fbank80}+win25+concat+resnet & 18.20                 \\
    ~    & fbank80+\textbf{win15}+concat+resnet & 17.68                  \\
    ~    & fbank80+win15+\textbf{GLU}+resnet & 17.12                  \\
    ~    & fbank80+win15+GLU+\textbf{mobilenet} & 16.92               \\
    \hline
    \multirow{5}*{conformer}     & fbank26+win25+concat+resnet & 16.12                    \\
    ~    & \textbf{fbank80}+win25+concat+resnet & 15.64                 \\
    ~    & fbank80+\textbf{win15}+concat+resnet & 15.41                \\
    ~    & fbank80+win15+\textbf{GLU}+resnet & 14.25                 \\
    ~    & fbank80+win15+GLU+\textbf{mobilenet} & 14.26                 \\

    \hline
  \end{tabular}
\end{table}

\begin{table}[t]\footnotesize
\centering
  \caption{\noindent{Comparison of models, which is pre-training on CSTS for the first phase and fine-tuning on MISP, with different noise type (CER:\%).}}
  \label{tab:word_styles2}
  \begin{tabular}{|c|c|c|c|c|}
     \hline
 \multirow{2}*{\textbf{Model}} & \multirow{2}*{\textbf{Mode}} & \multicolumn{2}{|c|}{\textbf{SNR=5}}&\multirow{2}*{\textbf{clean}}  \\
 \cline{3-4}
  ~ & ~ &  \ \textbf{All} & \textbf{Babble}&~  \\
 \hline
AV-HuBERT   & A  & 92.40 & 116.75 & 34.56  \\
AV-HuBERT   & AV   & 45.34 & 54.80& 18.93  \\
conformer enhanced AV-HuBERT   & A  & 55.19 & 74.78  & 17.25 \\
conformer enhanced AV-HuBERT   & AV   & 30.48 & 38.40& 14.26 \\
 \hline
  \end{tabular}
\end{table}

\begin{table}[t]\footnotesize
  \caption{\noindent{Comparison between WeNet, AV-HuBERT, and enhanced AV-HuBERT. The last two models are pre-training for five phases and fine-tuning on MISP and CMLR (CER:\%).}  }
  \label{tab:word_styles2}
  \centering
  \begin{tabular}{|cc|c|c|c|}
    \hline
    \textbf{Model} & \textbf{Backbone} & \textbf{Labeled} &  \textbf{Unlabeled} & \textbf{CER} \\
    \hline

    {WeNet}    & {conformer} & 100hrs & - & 15.04                 \\
    AV-HuBERT    & transformer & 100hrs & 1000hrs & 12.95                  \\
    enhanced AV-HuBERT    & transformer & 100hrs & 1000hrs & 12.05                  \\
    \hline
    {WeNet}    & {conformer} & 61hrs & - & 3.65                  \\
    AV-HuBERT   & transformer & 61hrs & 1000hrs & 3                  \\
    enhanced AV-HuBERT    & transformer & 61hrs & 1000hrs & 2.82                  \\

    \hline
  \end{tabular}
\end{table}

In Table 3, we set the update frequency to 4, and leads to 4 times epoch growth. Compared with Table 2, the baseline WER in audio-only and audio-visual cases drops obviously. At the same time, we find that by incorporating conformer and all other boosting methods, WER can decrease around 7-16\% except NWER in the audio-only experiment. 

In Table 4, we conduct a series of experiments on the transformer and conformer backbone group, where 4 methods are presented for each backbone, including filterbank dimension, frame length, visual extractor category, and audio-visual fusion type. Our baseline sets filterbank to 26-dim, and widow size to 25ms. Only one factor is compared to each line in each backbone experiment. For the reader's convenience, we present our results in an increasing way of performance. Conclusions are:

1) Filterbank with a higher dimension is relatively better(80-dim vs 26-dim) with 3.8\% relative CER improvement. Increasing the number of filters when extracting audio features leads to more sufficient knowledge and thus better feature representation.

2) Decrease frame length boosts performance, where 15ms outperforms 25ms window by 2.8\% relative CER improvement. One unanticipated finding was that our outcome is contrary to the instinct longer frame length was considered to provide better feature representation where more information lies in there.

3) Gated fusion purposefully accepts audio-visual information flow, which exceeds vanilla concatenation by 3\%-7\% on relative CER performance, from transformer and conformer experiments. 

4) Mobilenet improves lightly and we think the main reason is that mobilenet has 7M fewer parameters than resnet18.

In Table 5, we compare the robustness of the conformer-enhanced AV-HuBERT and AV-HuBERT under various noise types with a 5dB signal-to-noise ratio (SNR) in the first phase. All means noise will be selected randomly from babble, music, and speech noise. The conformer-enhanced AV-HuBERT uses an 80-dim fbank, 15ms window size, gated fusion, and MobileNet. It can be seen that the conformer-enhanced AV-HuBERT outperforms AV-HuBERT no matter in clean or different noisy conditions. AV is better than A by adding visual modality. The experimental results show that our proposed conformer-enhanced AV-HuBERT does have a significant improvement over AV-HuBERT.

In Table 6, we compare the performance of the AV-HuBERT baseline model, the enhanced AV-HuBERT model, and the single audio modality-trained WeNet model.
It can be seen that, due to extra modality in AV-HuBERT and 1000h data for pre-training, AV-HuBERT surpasses WeNet CER by relatively 14\% and 18\% on MISP and CMLR. Moreover, the proposed enhanced AV-HuBERT outperforms the baseline AV-HuBERT by 7\% and 6\%, which shows the effectiveness of our proposed boosting methods and the generalization of the CSTS dataset we established.

Looking into CER results, performance on CMLR is admirably up to 2.82\%, compared with that of MISP, which is 12.05\%. The logic behind this could be ascended to the nature of the dataset. More specifically, CMLR contains clear News reports and high-quality audio leading to better performance while MISP is relatively more challenging, with HeFei accent chatting data, background noise sometimes, and blur visual resources.




\section{CONCLUSION \& FUTURE WORK}
\label{sec:majhead}
This paper dives into the practice of multi-modality pre-training framework on Mandarin and English datasets. We establish a 1000h Mandarin multi-modality dataset, CSTS. With the help of CSTS, we verify that AV-HuBERT with one extra modality and pre-training stage can outperform WeNet with one modality. CER drops 14\% and 18\% relatively on the Mandarin dataset MISP and CMLR. Based on AV-HuBERT, we propose the conformer enhanced AV-HuBERT, which also surpasses the baseline by a relative 7\% and 6\% in CER.  
Further work will focus on:

\noindent\textbf{Attention-based AV alignment.}
Attention-based approaches are used in \cite{Chang2021MultiChannelTT} to automatically align audio with video. Referring to the methods in the paper above, self-attention will perform over audio and video respectively. Then, the matrix containing video information will play the role of key and value while the audio matrix will be treated as the query. Finally, they will learn the alignment through the cross-attention approach.

\noindent\textbf{Trainable audio convolution network.}
A trainable audio convolution network has the potential to take the place of a filterbank to extract audio features. Experiments from google \cite{Sainath2015LearningTS} also indicate a decent adjustment.

\noindent\textbf{AV-Confidence.}
\cite{Yu2021FusingIS} suggests the usage of audio-visual confidence, besides current features from lip movement and speech. A decision fusion net combines all modality information, including confidence, which would help our AVSR system we believe.

\bibliographystyle{IEEEbib}
\bibliography{refs}

\end{document}